# Generation of ultrashort (~10ps) spontaneous emission pulses by quantum dots in a switched optical microcavity

E. Peinke,[1] T. Sattler,[1] G. M. Torelly,[2] J. Bleuse,[1] J. Claudon,[1] W. L. Vos,[3] J. M. Gérard[1,a]

[1]*Univ. Grenoble Alpes, CEA, IRIG-PHELIQS,"Nanophysique et semiconducteurs" group, F-38000 Grenoble, France*
[2] *LabSem-CETUC, Pontifícia Universidade Católica do Rio de Janeiro, 22451-900, Brazil*
[3]*Twente University, MESA+, COPS team, Enschede, The Netherlands*
(Dated: October 2, 2019)

We report on the generation of few-ps long spontaneous emission pulses by quantum dots (QDs) in a switched optical microcavity. We use a pulsed optical injection of free charge carriers to induce a large frequency shift of the fundamental mode of a GaAs/AlAs micropillar. We track in real time by time-resolved photoluminescence its fundamental mode during its relaxation, using the emission of the QD ensemble as a broadband internal light source. Sub-ensembles of QDs emitting at a given frequency, interact transiently with the mode and emit an ultrashort spontaneous emission pulse into it. By playing with switching parameters and with the emission frequency of the QDs, selected by spectral filtering, pulse durations ranging from 300 ps down to 6 ps have been obtained. These pulses display a very small coherence length, which opens potential applications in the field of ultrafast imaging. The control of QD-mode coupling on ps-time scales establishes also cavity switching as a key resource for quantum photonics.

Over the last 25 years, optical microcavities have been widely used to tailor the optical properties of semiconductor emitters in solid-state cavity quantum electrodynamics (CQED) experiments [1-5]. The strong coupling regime has been observed for quantum wells in planar cavities [6], as well as for a single quantum dot (QD) coupled to a single cavity mode [5]. In the weak coupling regime, the enhancement [1,3,7] and the inhibition [2-8] of the spontaneous emission (SpE) rate have both been observed for QDs. In these experiments, the coupling between the emitter and the electromagnetic field is drastically different from that of the "emitter in free space" case. However, this coupling is kept constant for the entire duration of the SpE event, in stark contrast with atomic CQED experiments, for which the interaction time between a flying atom and a cavity mode can be tuned by playing with the atom velocity [9].

We have recently introduced a new paradigm in solid-state CQED [10] by proposing to modify the emitter-field coupling on a time scale shorter than the emitter lifetime, through an ultrafast control of cavity modes' frequencies. Reversible frequency switching of cavity modes was achieved on a sub-picosecond time scale using the electronic Kerr effect [11,12], and within a few ps through optical injection of free carriers [13-18]. This latter scheme is very well suited for the dynamic control of QD-cavity systems in the weak coupling regime, since the free carriers lifetime ranges between a few tens of ps and a few ns. A reversible change of the SpE rate of a QD in a photonic crystal on a 200-ps time scale was recently demonstrated using this approach [18]. We report in this paper on the generation of ultrashort (a few ps long) SpE pulses without temporal coherence by QDs in a switched micropillar cavity. Such pulses are emitted by the QDs during a transient coupling event, induced by the ultrafast switching of the cavity modes by optically generated free carriers.

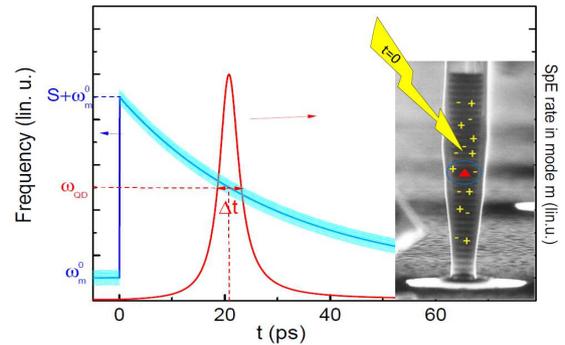

FIG.1. Concept for the generation of short SpE pulses. Right: a micropillar mode is switched at t=0 using free carrier injection. Left: mode frequency $\omega_m$ (blue line, broadened by the mode linewidth $\Delta\omega_m$) and intensity of QD SpE in the mode (red line) as a function of time. The transient QD-mode coupling leads to the emission of a few ps long SpE pulse. The red curve has been calculated as in [5], using typical experimental parameters (see text).

We sketch in Fig. 1 the experimental approach which has been used for this demonstration. We consider firstly an optical microcavity, such as a GaAs/AlAs micropillar cavity, containing an InAs QD. The QD bandgap is at the frequency $\omega_{QD}$, which belongs to the frequency interval $[\omega_m^0, \omega_m^0 + S]$ that is swept by mode $m$ during the switching events induced by free carrier injection. Here, $S$ is the frequency switch amplitude. At t=0+, a pump pulse excites the QD-cavity system, shifting the mode to $\omega_m^0 + S$ and exciting the QD. The QD does not emit light into mode $m$, because of the large QD-mode detuning. The mode relaxes then down to its original frequency $\omega_m^0$, entering eventually into resonance with the QD during a short time $\Delta t = \Delta\omega_m/(d\omega_m/dt)$, where $\Delta\omega_m$ is the mode linewidth and $d\omega_m/dt$ the shifting speed of the mode in the frequency domain. Thanks to the Purcell effect, a large fraction β of the QD SpE is injected into the cavity mode during this transient coupling event [1], after which the QD emission is again detuned with respect to mode $m$. We therefore expect the QD to emit a short pulse of light of duration $\Delta t$ into mode $m$ as described theoretically in [10]. As shown later, based on a characterization of micropillar switching events, typical values are 0.4 meV and 0.04 meV/ps for $\hbar\Delta\omega_m$ and $\hbar\, d\omega_m/dt$ respectively, which leads to $\Delta t \sim 10$ps. Therefore, we conclude that switched micropillars containing a single QD can be used to generate a few-ps-long SpE pulse.

---
[a] jean-michel.gerard@cea.fr

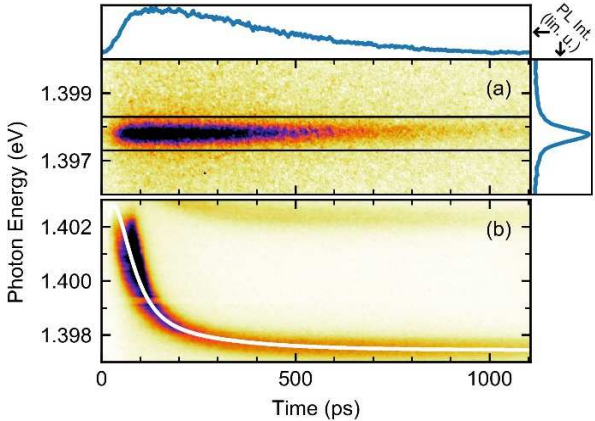

FIG. 2. (a) Streak camera image, PL spectrum (top curve) and PL decay profile (right curve) obtained for a QD-micropillar system for $E_{pump}$ =1.47 eV. (b) Streak camera image obtained for $E_{pump}$ =1.55 eV. The white line corresponds to a biexponential fit of the mode relaxation. The 3-ps pump pulse is centered at t=0ps.

Short pulse generation is in fact also possible for micropillars containing an ensemble of QDs, despite the inhomogeneous distribution of their bandgaps, provided a spectral selection of QDs emitting in a narrow spectral window is implemented. QDs emission into the mode will then occur, as long as the cavity mode and the collection window frequencies overlap. The duration of the SpE pulse is then simply given by:

$$\Delta t = \sqrt{\Delta\omega_m^2 + w^2} \Big/ \frac{d\omega_m}{dt}(t_{res}) \quad (1)$$

where $w$ is the width of the frequency window, and $t_{res}$ the time at which the mode frequency $\omega_m$ corresponds to the central frequency $\omega_w$ of the collection window.

To demonstrate this effect, we investigate GaAs/AlAs micropillar cavities which contain a large collection of InAs QDs (~20000) in the GaAs cavity layer, which is surrounded by two Bragg mirrors, comprising 25 (bottom side) and 15 (top side) pairs of GaAs/AlAs layers. The growth conditions are chosen to ensure a relatively high emission frequency of the QD ensemble (~1.38 eV) at 8K [19]. The QDs are located close to the axial antinodes of the cavity modes to ensure optimal coupling. The detailed structure and fabrication process of these cavities are described in the supplemental material (SM [20]).

Cavity switching events are usually studied using pump-probe optical spectroscopy [8-10]. We use here a different approach, inspired by early work on static microcavities [21]. We use the QD ensemble as a broadband internal light source, which feeds the cavity modes. Switching events are probed on a ps time scale with a time-resolved microphotoluminescence (PL) setup, described in [20]. The QD-micropillar system is excited by a Ti:sapphire laser delivering 3-ps pulses at a 76 MHz repetition rate. A microscope objective is used to focus the laser beam and to collect the PL. The laser beam waist is slightly broader than the pillar to ensure a good uniformity of the density of photogenerated charge carriers. The setup combines a grating spectrometer and a streak camera to analyze both spectrally and temporally the emission of the QD-micropillar system and to track, in a dynamic way, the resonant cavity modes during switching events. Sharp pass-band filters or the spectrometer itself are used to select, when needed, the QDs' emission within a chosen spectral window and to generate short spontaneous emission (SpE) bursts.

We first use this versatile setup to probe the optical properties and the switching characteristics of our QD-microcavity system at 8K. Figure 2 shows typical emission results obtained for QDs in a 3-µm diameter GaAs/AlAs micropillar. In a first experiment, the incident laser photon energy ($E_{pump}$=1.47 eV) ensures an efficient excitation of the QDs by generating electron-hole pairs in the InAs wetting layer, while avoiding absorption in bulk GaAs. The pulse intensity (1 pJ) is adjusted so as to prevent saturation of the QD emission. As shown in Fig. 2a, this experiment enables one to determine from the PL spectrum the frequencies $\omega_m^0$ and quality factors of the first resonant modes of the unswitched micropillar. We use in the following the standard notation $HE_{11}$ for the fundamental mode. Considering now QD properties, we observe a modest twofold enhancement of the QD SpE rate (QD exciton lifetime $\tau_{QD}$ around 530 ps instead of 1.1 ns in bulk GaAs) for QDs coupled to mode $HE_{11}$, as expected when considering the small Purcell factor $F_p$ of this mode ($Q$= 3000, $F_p\sim6$) and the averaging of the Purcell enhancement for QD ensembles [1,7].

In a second experiment, we tune the Ti:sapphire laser energy to $E_{pump}$ =1.55 eV, in order to inject electron-hole pairs in the GaAs layers. We chose on purpose $E_{pump}$ slightly above the bandgap of GaAs (1.52 eV). Since the density of accessible states for electron-hole pairs in GaAs is small, the pump propagates without significant depletion. This ensures a uniform change of the refractive index of the GaAs layers over the entire structure. For these pumping conditions ($E_{pump}$=1.55 eV, 30 pJ/pulse), the density of electron-hole pairs created by the pump pulse arrival is of the order of $3 \times 10^{17}$ cm$^{-3}$. Some of the free carriers are also captured by the QDs, the broadband emission of which can be used to probe the time-dependent photonic mode density of the micropillar.

As soon as the QD photoluminescence (PL) is detectable (i.e. after a 20-ps delay due to carrier capture and relaxation into the QDs), we observe in Fig. 2b mode $HE_{11}$ at a higher frequency compared to that of Fig. 2a (first experiment). The blueshift is around 5.5 meV, which is as large as 12 mode linewidths. This fast "switch-on" agrees with data obtained in pump-probe experiments (7 ps in [16, 22]). We observe next a slower relaxation of the mode towards its initial frequency, due to the recombination of electron hole-pairs. For long delays (>200ps), the decay is mostly governed by carrier diffusion and non-radiative recombination at etched pillar sidewalls [13, 14, 22]. Exponential relaxation with a 200ps time constant is observed over this time range. At shorter delays after the pulse, the high density electron-hole plasma experiences also an efficient radiative recombination, leading to a much faster relaxation. Overall, the time dependence of the mode frequency $\omega_l(t)$ is well approximated by a biexponential decay with characteristic time constants 40ps and 200ps, as shown in Fig. 2b.

From the frequency shift S=5.5 meV, we can infer a relative index change $\Delta n/n \sim 3.6 \cdot 10^{-3}$ for GaAs, in agreement with previously reported work [22]. Free carriers are also expected to induce absorption losses ($\approx$ 50 cm$^{-1}$ for $2 \cdot 10^{17}$ pairs/cm$^3$ [23]). However, these extra losses are small compared to intrinsic losses, so that we do not observe any sizeable degradation of Q for switched modes, pointing out that switched QD-micropillars (SQMs) are well suited for dynamic CQED experiments.

As a first example of SpE control on a few-ps timescale, we demonstrate in Fig. 3 the generation of light bursts with adjustable durations. We use the spectrometer to select the emission from a sub-ensemble of QDs, emitting within a 0.3-meV wide frequency window. We plot in Fig. 3a the temporal profiles of the light pulses obtained for two different frequency windows. For windows W1 and W2, we observe a single pulse, related to the resonance of the QDs with relaxing mode $HE_{11}$. As suggested by Eq.1, its duration can be tuned by playing with the detuning of the frequency window $\omega_w$ with respect to $\omega_1^0$ since the shifting speed $d\omega_l/dt$ increases with this detuning. As an example, the detuning $\omega_w - \omega_1^0$ is equal to 1 (5) meV for W1 (W2), leading to burst durations (defined as the pulse width at half-maximum) equal to 200ps for W1 and as small as 6ps for W2.

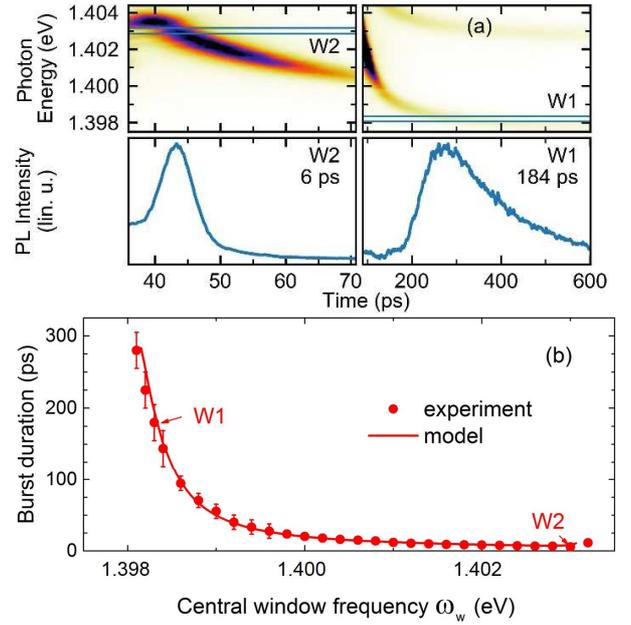

FIG. 3. (a) Top: Spectrally and temporally resolved emission for the two first modes of the SQM under strong pumping conditions ($E_{pump}$=1.55 eV, P=30pJ/pulse). Two different time ranges are used for sake of clarity. Collection windows W1 and W2 are marked by solid lines (Bottom) Temporal analysis of the SQM emission for W1 and W2. (b) Measurements (dots) and theoretical estimates (lines) of the duration of SpE pulses emitted in mode $HE_{11}$, as a function of the central window frequency $\omega_w$ (w=0.3 meV).

We plot in Fig. 3(b) pulse durations deduced from experiments, which cover a wide time range by only playing with the central window frequency $\omega_w$. A modeling of burst durations based on the knowledge of $\omega_l(t)$ and Eq.1 accounts very well for these experimental results. Short SpE pulses can of course also be emitted by QD ensembles in static cavities in the Purcell regime. However, their duration cannot be shorter than around 30-ps, even for giant Purcell factors ($F_p$> 100), because of the time jitter induced by carrier capture and relaxation times under non-resonant pumping conditions [1, 24]. Cavity switching appears as an utmost new approach to generate short pulses of light in a controlled way.

Other devices based on micropillars, the vertical cavity amplifying photonic switches [25], have been used to prepare ~20-ps long light pulses using a pump-probe experimental scheme. In these devices, a probe beam is amplified in a transient way, by the stimulated emission from a quantum well that is excited by a pulsed pump. By contrast, we expect only SpE to be at work in experiments shown in Fig. 3. Because $\omega_w$ is located within the high-energy tail of the QD ensemble [19], the number N of QDs contributing to emission pulses with their fundamental optical transition is smaller than 100. Therefore, a simple estimate shows that the average number of photons in the cavity mode, given by $N.\beta.\tau_{cav}/\tau_{QD}$ (where $\tau_{cav}$ is the storage time for mode *m*), does not exceed 0.5. As a result, stimulated emission can be *a priori* excluded.

In order to bring experimental support to this claim, we have probed the temporal coherence of the pulses produced by the SQMs by studying their transmission through a 500-µm Teflon film, used as scattering medium. At around 1.4 eV, the photon mean free path is around 50 µm [26], i.e. much smaller than the thickness of our film, so that light experiences many scattering events before escaping at the exit side. For a large enough coherence length of the impinging beam, i.e. larger than the optical length difference for paths resulting from different scattering events, one expects the formation of a 2D speckle pattern at the output facet of the diffusor. For this experiment, we use an imaging system based on a Si charge-coupled device (CCD) camera to probe the appearance of speckles. As a reference source, we use 3-ps long pulses at $\omega_{pump}$ =1.4 eV. Since $L_{coh}$ is much larger than the film thickness for laser pulses, a clear speckle pattern is observed (see Fig.4. and the SM). This pattern is identical for two independent acquisitions, and does not depend upon pulse energies, as expected.

A drastically different behavior is observed for SQM pulses, as shown in Fig.4 for a 2-µm diameter micropillar that emits 18ps SpE bursts. We use mode $HE_{11}$ for its highly directive emission pattern [1, 3] and optical filters to define the collection window ($\omega_w$ =1.407 eV, *w*= 3 meV). The SQM pulses are well above the Fourier limit ($w.\Delta t \approx$ 100 »1). Assuming they stem from the sum of uncorrelated SpE from numerous QDs spread all over the collection window, one expects a coherence time $\tau_{coh}^{ref}$ smaller than $1/w \approx 0.12$ps and a coherence length of about 35 µm, much smaller than the thickness of the Teflon film [26]. As shown in Fig.4, no speckles are observed experimentally for SQM pulses under weak excitation conditions (P=60pJ/pulse). Intensity fluctuations between CCD pixels do not exceed the expected level for random detection events, and images obtained from different acquisition runs are uncorrelated, as shown in more detail in the SM. Under stronger pumping conditions (P> 80 pJ/pulse), this QD-micropillar does lase due to the filling of QD ground states and extra gain provided by upper optical transitions. As shown in Fig.4 and in the SM, the thresholds for lasing and for speckle formation are nicely in line: coherence-free pulses are only generated by SQMs in the SpE regime, a crucial point for potential applications.

The interest of low-coherence light pulses for imaging is well recognized, as they enable e.g. detection of objects hidden by a turbid environment [28] or obtaining clear images of ultrafast processes in fluids [29]. Laser pulses are indeed poorly adapted to such applications, as their coherence induces diffraction fringes and speckle patterns in recorded images. State-of-the-art low-coherence sources exploit few-ns long fluorescence pulses emitted by a laser dye [29]. SQMs could in principle be used to extend ultrafast imaging to the sub-ns range, which is of interest for various domains (fluidics, shock wave propagation among others). At first sight, SQMs suffer from their small power. Their mere operation principle limits the average photon number per pulse to $\Delta \tau/\tau_{cav}$ so as to avoid stimulated emission, e.g. 10 photons/pulse for $Q$=1500 and $\Delta t$=10ps. By using longer pulses ($\Delta t \sim 200$ps) and operating a large array (100 x 100) of very similar SQMs in parallel, one would reach $2.10^6$ photons/pulse. Though challenging, this strategy is likely to open a route towards the imaging in a stroboscopic mode of highly reproducible processes such as inkjet droplet formation [29] on a sub-ns time-scale. Thanks to its high repetition rate (76 MHz here, with potential increase up to 5GHz), a single SQM could deliver up to $10^{11}$ photons/second. Therefore, the detection of objects hidden in turbid media such as fog or murky water [28] can be considered a realistic target in the short term.

In the context of quantum photonics, this work provides a first example of QD SpE switching of QD SpE in a cavity mode on a few-ps time scale. This could be used to switch on/off the strong coupling regime between a single QD and a single cavity mode. Temporal shaping of pump pulses could also be used to play more smoothly on the QD-mode detuning. When applied to state-of-the-art micropillars with Purcell factors $F_p$ around 400 [30], such a smooth tuning would enable to control the magnitude of the Purcell effect in real-time and to engineer the time-envelope of single photon pulses emitted by a QD as modelled in [31], e.g. to enable perfect reabsorption by a different QD in quantum networks [32].

The authors gratefully acknowledge fruitful discussions with P.L. Souza and B. Gayral. This work was supported by ANR Project NOMOS (ANR-16-CE09-0010-01). G.M.T. acknowledges funding from CAPES (PDSE 88881.187056/2018.1) and CNPq.

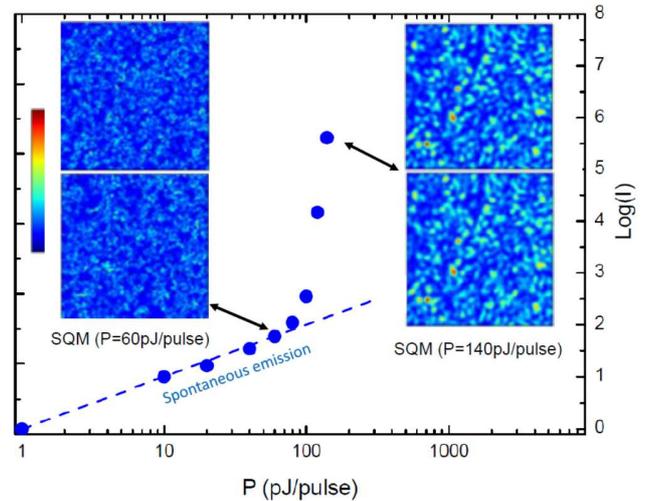

Fig. 4. Evolution of the intensity *I* of the pulses delivered by the switched QD-micropillar (blue dots) as a function of the average pumping power P of the Ti-sapphire laser, plotted in log-log scale. For powers *P* below ≈ 70 pJ/pulse, it follows the linear trend expected for pure spontaneous emission, shown as a dotted blue line. For larger pump powers, the onset of stimulated emission of the QD ensemble results in a superlinear increase of *I*. The insets show CCD camera images of the output facet of a Teflon film, upon irradiation by SQM pulses, for two different pulse energies. For each energy, we show two images obtained from independent acquisitions. Speckle formation is clearly observed in the strong pumping regime.

# Supplemental Material for "Generation of ultrashort (~10ps) spontaneous emission pulses by quantum dots in a switched optical microcavity"

E. Peinke[2], T. Sattler[1a], G.M. Torelly[2], J. Bleuse[1], J. Claudon[1], W.L. Vos[3] and J.M. Gérard[1]

[1]Univ. Grenoble-Alpes, CEA, INAC/PHELIQS/ CEA-CNRS Nanophysics and Semiconductors joint team
38000 Grenoble, France

[2] CETUC-LabSem, Pontifícia Universidade Católica do Rio de Janeiro, 22451-900, Brazil

---

[2] E. Peinke and T. Sattler contributed equally to this work.

*³Twente University, MESA+, COPS team, Enschede, The Netherlands*
*e-mail: jean-michel.gerard@cea.fr*

**Sample fabrication**

Micropillars containing quantum dots (QDs) have been processed from a planar microcavity M, grown by molecular beam epitaxy. M is formed by a GaAs one-wavelength thick cavity layer, surrounded by a 15 (resp. 25) period distributed Bragg reflectors (DBRs) made of quarter-wavelengths GaAs and AlAs sublayers on the top side (resp. bottom side). This planar microcavity is designed for an operation wavelength around $\lambda$=0.96 µm at a temperature of 4K, assuming refractive indices equal to 3.5 for GaAs and 2.95 for AlAs.

Five layers of InAs QDs (areal density ~4. $10^{10}$ cm$^{-2}$ per layer) have been inserted within the cavity layer, close to the antinodes of the resonant mode in the planar cavity. More precisely, two layers are located at 10 nm from the interface with the top or bottom DBR's, and the other three are placed at the center of the cavity layer, 10 nm above the center, and 10 nm below the center. In order to form the InAs QD layers, we deposit 0.6 nm of InAs within 1 s at 520 °C, and the GaAs overlayer is immediately grown on top at the end of the deposition of InAs. With this procedure, small QDs are nucleated, not significantly perturbing the subsequent growth of the top DBR, and emitting at a relatively large energy (which is interesting to better match the response window of our detection system). In a reference sample without DBR's, the emission of such QD ensembles is spectrally broad (60 meV) due to QD size fluctuations, and centered around 1.36 eV [SM1].

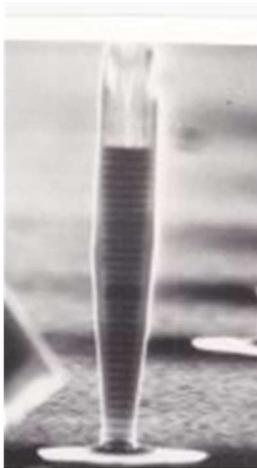

Fig.S1 : Scanning electron micrograph of a QD-micropillar used for this study. The diameter of the top facet is 1 µm.

Pillar microcavities, known as micropillars, have been processed from microcavity *M* according to the following procedure [SM2]. A 2 µm thick ''hard-mask'' layer consisting of backed optical resist $Si_3N_4$ is first deposited on the sample. Electron-beam lithography using polymethylmethacrylate and the lift-off technique are used to define a 100 nm thick Al mask, which is transferred to the hard-mask layer by reactive ion etching (RIE) using a $SF_6$ plasma. RIE using $SiCl_4$ is then performed to etch the epitaxial structure. As shown by the scanning electron micrograph shown in figure S1 for a 1 µm diameter micropillar, straight and smooth sidewalls are obtained this way.

**Experimental set-up**

The QD-micropillar sample is placed at a temperature around 5 K inside a helium-flow cryostat. A single micropillar at a time is excited by a Ti:sapphire laser delivering 2.9 ps pulses at a 76 MHz repetition rate. We use an epi-microfluorescence experimental configuration, for which the same microscope objective (Zeiss 441030-9901, NA=0.25) is used to focus the laser beam, and to collect the QD-micropillar emission, which is shaped into a collimated beam.

For the characterization of the light pulses emitted by switched QD-micropillars (SQMs), we use a system composed of a Jobin-Yvon Triax320 monochromator and a Hamamatsu C10910 streak camera. This system provides a temporal resolution around 2 ps, combined to a 0.3 meV resolution in the photon energy/frequency domain, close to the limit imposed by the time-energy uncertainty principle.

For the test of the temporal coherence of SQM pulses, we select the emission of QDs within a chosen frequency window around 1.4 eV using pass-band filters (Semrock LP02-830RU-25). The collimated beam impinges on a 0.5-mm-thick film of poly-tetrafluoroethylene (PTFE) (Goodfellow, ref. FP301400) under normal incidence. We image the opposite facet of the film on a CCD camera (model Coolsnap-ES from Roper Scientific).

**Preparation and characterization of SQM pulses**

We

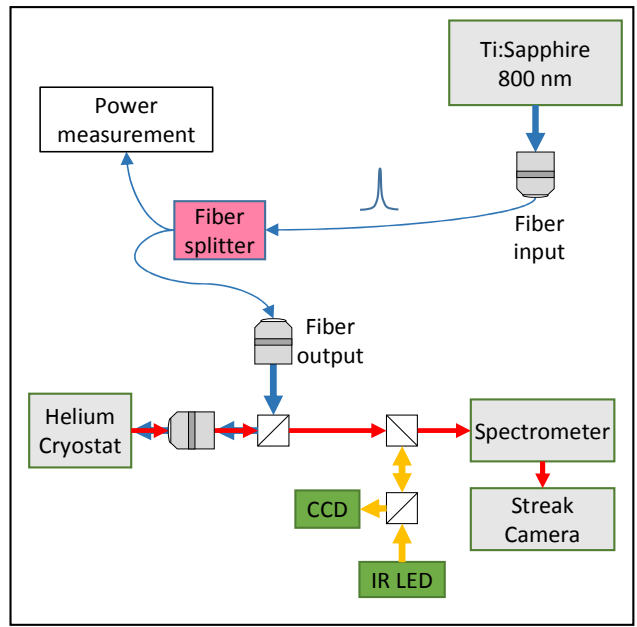

Fig. S2 Schematic view of the experimental set-up

describe in this section the properties of the SQM pulses which have been used for a qualitative test of their temporal coherence. Here, the micropillar has a 2 µm diameter and the frequency of its (unswitched) fundamental mode $HE_{11}$ is 1.402 eV. The filters are used to define a 3 meV broad collection window [1.4055 eV-1.4085 eV]. This SQM exhibits a large switching amplitude ($S$> 7 meV for $P$> 1 mW) and fast return dynamics due to efficient non-radiative recombination at sidewalls ($d\omega_m/dt$ = 0.16 meV/ps). We use the optical filters to define [1.4055-1.4085 eV] the collection window. The laser pump is at 1.7 eV, and we vary the pump power $P$ from 10 pJ/pulse to 140 pJ/pulse.

We first analyze with our monochromator/streak camera system the light pulses which are prepared this way. As shown in figure S3, their duration is 18 ps, in agreement with equation (1) of the main manuscript. At this stage, it is important to note that the switching amplitude is always larger than 7 meV when we vary the pump power $P$ from 10 to 140 pJ/pulse. Therefore, the mode $HE_{11}$ spans in all cases the entire collection window and the duration of the emitted pulse is unchanged, and equal to 18ps for all pump powers.

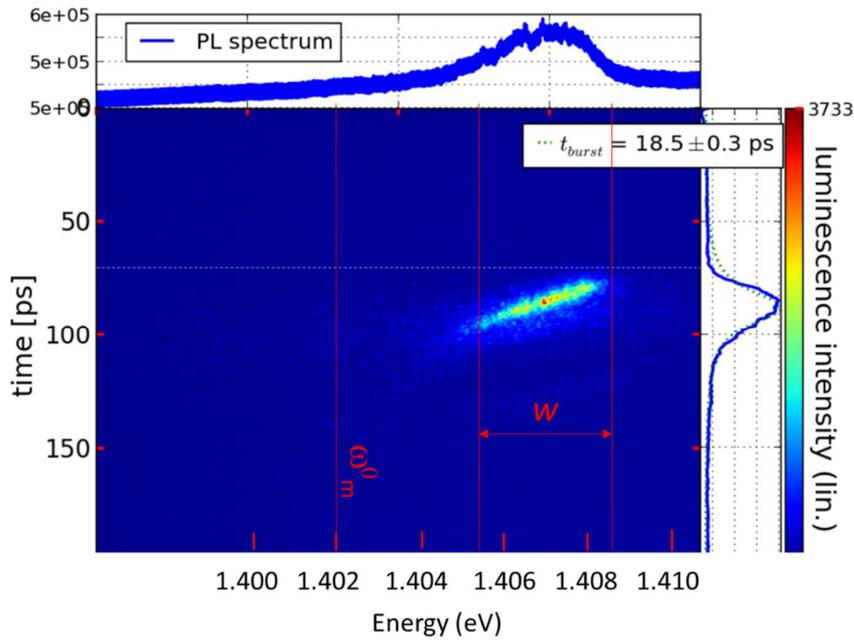

Figure S3: Temporal and spectral analysis of pulses emitted by a SQM (2 µm pillar radius, $\omega_0$ =1.402 eV, S= 7 meV, [1.4055 eV-1.4085 eV] collection frequency window).

We plot in Figure 4 of the main manuscript, the dependence of the intensity *I* of SQM pulses as a function of the pump power. For low powers, *I* follows a linear law, as expected for a pure spontaneous emission process. However, a clear super-linear behavior is observed above 80 pJ/pulse. For these SQM, stimulated emission and lasing are in fact observed for high pump powers, due to filling of fundamental QD states and to additional gain provided by optical transitions between excited QD states.

**Test of the temporal coherence of SQM pulses**

To test the temporal coherence of SQM pulses, the collimated beam of the SQM impinges the surface of the 0.5 mm thick PTFE film over a 1mm x 1mm area. We image the output facet of the PTFE film with a 1000 x 1000 pixel CCD sensor. The images that are inserted in Figure S4 correspond to a 80 x 80 pixels zone of the CCD sensor, selected near the center of the image and corresponding to a 50 µm x 50 µm area of the PTFE film.

In a first experiment, we use a reference laser beam, delivered by the pulsed Ti-sapphire laser. Its femtosecond laser pulses are stretched into picosecond pulses (separated by 13 ns) after propagation through a ≈1 m long optical fiber. Their shape is very close to a Gaussian and their full-width at half maximum is $\Delta t$ = 2.9 ps. Since pulses go through the PTFE film one by one, the relevant coherence length $L_{coh}^{ref}$ is approximately given by $L_{coh}^{ref}$ = c x $\Delta t$ = 900 µm. As shown in figure S4, we observe in this case a clear speckle pattern at the output facet of the PTFE film. This is well understood, since the coherence length of the light pulses is larger than the optical thickness of the PTFE film.

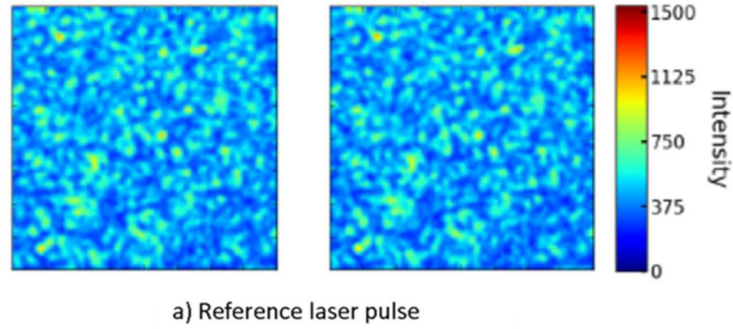

a) Reference laser pulse

Fig. S4. 80 x 80 pixels intensity maps measured at the output facet of a 0.5 mm thick PTFE film upon transmission of 3 ps -long reference laser pulses delivered by the mode-locked Ti-sapphire laser at around 1.4 eV. Maps are shown for two consecutive acquisitions. A speckle pattern is clearly observable.

We now consider the results obtained with the light pulses that are generated by our SQM. In the spontaneous emission regime ($P_{pump}$ <80 pJ/pulse), we obtain noisy, but uniformly bright images as shown in the Fig. 4 of the main manuscript for $P_{pump}$ = 60 pJ/pulse. More precisely, denoting $C_{nm}$ the number of counts for pixel (n,m) and $\bar{C}$ the average number of counts, the standard deviation σ, which reflects intra-image pixel intensity fluctuations is given by:

$$\sigma^2 = \sum_{nm}(C_{nm} - \bar{C})^2/\bar{C}^2$$

In this weak-pumping regime, σ corresponds to the shot noise level i.e. $\sigma/\sqrt{\bar{C}}$ ~1.

The absence of speckles can also be confirmed by comparing pairs of images obtained from two independent acquisition runs. We define a inter-image distance D, to estimate how different two images are from one another:

$$D^2 = \sum_{nm}\left(\frac{C_{nm}}{\bar{C}} - \frac{C'_{nm}}{\bar{C}'}\right)^2$$

where $C'_{nm}$ is the number of counts for pixel (n,m), and $\bar{C}'$ the average value of $C'_{nm}$ for the second image. Please note that we normalize the pixel intensity in this definition of the distance, to avoid spurious effects related to possible small drifts of the pump intensity between the two acquisitions.

Trivially, *D* is equal to zero for two identical images; more generally, for highly contrasted and strongly correlated speckled images (large σ, small *D*), one expects *D/σ* to become vanishingly small. By contrast, one can easily show that for unstructured images for which deviations of pixel intensities with respect to the average value are only due to random noise (no inter-pixel correlations), then $D/\sigma = \sqrt{2}$. This criterion is very nicely satisfied for the pair of images shown in the figure 4 of the main manuscript for P=60pJ/pulse.

Let us now consider the pairs of images obtained for $P_{pump}$= 140 pJ/pulse, which are also shown in the figure 4 of the main manuscript. "Hot spots" are clearly visible, and the two images look very similar. Speckle formation is easily evidenced by considering the two criteria which have just been introduced. On one hand, the pixel to pixel intensity fluctuation σ is, for both images, 250 times larger than the shot noise level. On the other hand, $D/\sigma = 0.12 \ll \sqrt{2}$, which highlights the high degree of similarity between these two highly contrasted images.

These results show that the coherence length of SQM pulses increases drastically when going from low to high power pumping regimes, due to the onset of stimulated emission. We plan to measure precisely in the future the coherence length $L_{coh}^{ref}$ of SQM pulses, and its dependence upon the pumping power using interferometric correlation spectroscopy [SM3], but this experiment goes beyond the scope of the present work.